\documentstyle[aps,pra,graphicx]{revtex}

\newcommand{\epsfigure}[3]{
 \begin{figure}
 \begin{center}
 \includegraphics[width=#3, keepaspectratio=true, clip=true]{#2}
 \nopagebreak
 \caption{#1}
 \end{center}
 \end{figure}}

\begin{document}

\twocolumn[\hsize\textwidth\columnwidth\hsize\csname
@twocolumnfalse\endcsname

\date{\today}

\title{Violation of Bell's inequality under strict Einstein locality
conditions}

\author{Gregor Weihs, Thomas Jennewein, Christoph Simon, Harald Weinfurter, and Anton Zeilinger}
\address{Institut f\"{u}r Experimentalphysik, Universit\"{a}t Innsbruck,\\
         Technikerstra{\ss}e 25, A--6020 Innsbruck, Austria}

\maketitle

\begin{abstract}
We observe strong violation of Bell's inequality in an Einstein,
Podolsky and Rosen type experiment with independent observers. Our
experiment definitely implements the ideas behind the well known
work by Aspect et al. We for the first time fully enforce the
condition of locality, a central assumption in the derivation of
Bell's theorem. The necessary space-like separation of the
observations is achieved by sufficient physical distance between
the measurement stations, by ultra-fast and random setting of the
analyzers, and by completely independent data registration.
\end{abstract}


\vskip1pc]

\narrowtext

The stronger-than-classical correlations between entangled quantum
systems, as first discovered by Einstein, Podolsky and Rosen (EPR)
in 1935\cite{Einstein35}, have ever since occupied a central
position in the discussions of the foundations of quantum
mechanics. After Bell's discovery\cite{Bell65} that EPR's
implication to explain the correlations using hidden parameters
would contradict the predictions of quantum physics, a number of
experimental tests have been
performed\cite{Freedman72,Aspect82,Kwiat95}. All recent
experiments confirm the predictions of quantum mechanics. Yet,
from a strictly logical point of view, they don't succeed in
ruling out a local realistic explanation completely, because of
two essential loopholes. The first loophole builds on the fact
that all experiments so far detect only a small subset of all
pairs created\cite{Pearle70}. It is therefore necessary to assume
that the pairs registered are a fair sample of all pairs emitted.
In principle this could be wrong and once the apparatus is
sufficiently refined the experimental observations will contradict
quantum mechanics. Yet we agree with John Bell that

\begin{quote}
"\ldots it is hard for me to believe that quantum mechanics works
so nicely for inefficient practical set-ups and is yet going to
fail badly when sufficient refinements are made. Of more
importance, in my opinion, is the complete absence of the vital
time factor in existing experiments. The analyzers are not rotated
during the flight of the particles."\cite{Bell87}
\end{quote}

This is the second loophole which so far has only been encountered
in an experiment by Aspect et al.\cite{Aspect82} where the
directions of polarization analysis were switched after the
photons left the source. Aspect et al., however, used periodic
sinusoidal switching, which is predictable into the future. Thus
communication slower than the speed of light, or even at the speed
of light \cite{Zeilinger86} could in principle explain the results
obtained. Therefore this second loophole is still open.

The assumption of locality in the derivation of Bell's theorem
requires that the measurement processes of the two observers are
space-like separated (Fig.~\ref{Spacetime}). This means that it is
necessary to freely choose a direction for analysis, to set the
analyzer and finally to register the particle such that it is
impossible for any information about these processes to travel via
any (possibly unknown) channel to the other observer before he, in
turn, finishes his measurement\cite{Weihs97}. Selection of an
analyzer direction has to be completely unpredictable which
necessitates a physical random number generator. A numerical
pseudo-random number generator can not be used, since its state at
any time is predetermined. Furthermore, to achieve complete
independence of both observers, one should avoid any common
context as would be conventional registration of coincidences as
in all previous experiments\cite{Christiansen85a}. Rather the
individual events should be registered on both sides completely
independently and compared only after the measurements are
finished. This requires independent and highly accurate time bases
on both sides.

In our experiment for the first time any mutual influence between
the two observations is excluded within the realm of Einstein
locality. To achieve this condition the observers ``Alice'' and
``Bob'' were spatially separated by 400~m across the Innsbruck
university science campus. We used polarization entangled photon
pairs which were sent to the observers through optical
fibers\cite{Tittel98}. About 250~m of each 500~m long cable was
laid out and the rest was left coiled at the source. This, we
remark, has no influence on the timing argument because the
optical elements of the source and the locally coiled fibers can
be seen as jointly forming the effective source of the experiment
(Fig.~\ref{Spacetime}). The difference in fiber length was less
than 1~m which means that the photons were registered
simultaneously within 5~ns.

As stated above, we define an individual measurement to last from
the first point in time which can influence the choice of the
analyzer setting until the final registration of the photon. In
order to rule out subluminal as well as luminal communication
between observers these individual measurements had to be shorter
than $1.3 {\rm \mu s}$, the time for direct communication at the
speed of light. This we achieved using high speed physical random
number generators and fast electro-optic modulators. Independent
data registration was performed by each observer having his own
time interval analyzer and atomic clock, synchronized only once
before each experiment cycle.

Our source of polarization entangled photon pairs is degenerate
type-II parametric down-conversion\cite{Kwiat95} where we pump a
BBO-crystal with 400~mW of 351~nm light from an Argon-ion-laser. A
telescope was used to narrow the UV-pump beam\cite{Monken98}, in
order to enhance the coupling of the 702~nm photons into the two
single-mode glass fibers. On the way to the fibers, the photons
passed a half-wave plate and the compensator crystals necessary to
compensate for in-crystal birefringence and to adjust the internal
phase $\varphi$ of the entangled state $ |\Psi\rangle =
1/\sqrt{2}(|H\rangle_1 |V\rangle_2 + {\rm e}^{i
\varphi}|V\rangle_1 |H\rangle_2)$, which we chose $\varphi=\pi$.

\epsfigure{Spacetime diagram of our Bell experiment. Selecting a
random analyzer direction, setting the analyzer and finally
detecting a photon constitute the measurement process. This
process on Alice's side must fully lie inside the shaded region
which is, during Bob's own measurement, invisible to him as a
matter of principle. For our setup this means that the decision
about the setting has to be made after point ``X'' if the
corresponding photons are detected at spacetime points ``Y'' and
``Z'' respectively. In our experiment the measurement process
(indicated by a short black bar) including the choice of a random
number only took less than a tenth of the maximum allowed time.
The vertical parts of the kinked photon world lines emerging from
the source represent the fiber coils at the source
location.\label{Spacetime}}{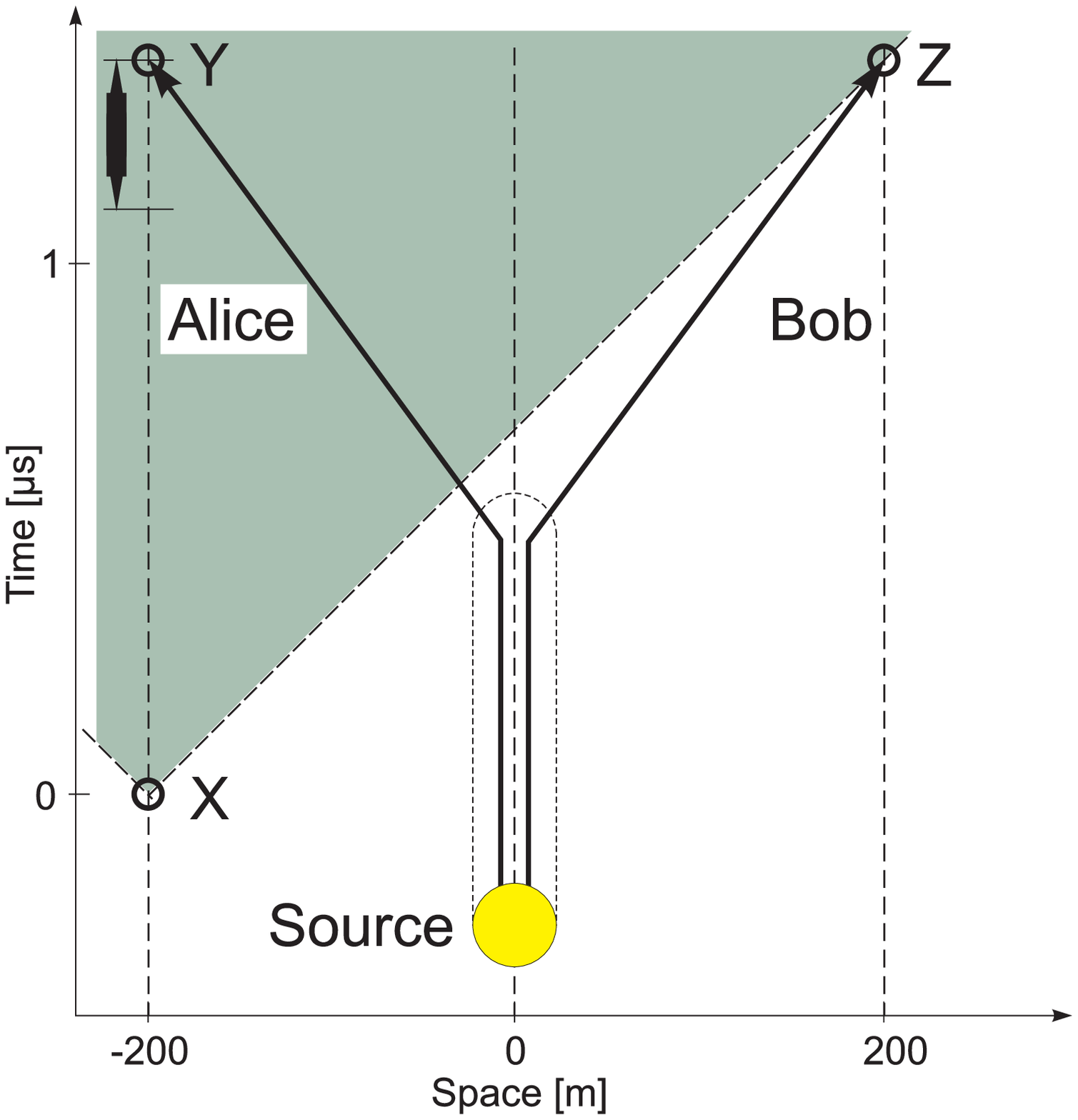}{0.7\columnwidth}

The single-mode optical fibers had been selected for a cutoff
wavelength close to 700~nm to minimize coupling losses. Manual
fiber polarization controllers were inserted at the source
location into both arms to be able to compensate for any unitary
polarization transformation in the fiber cable. Depolarization
within the fibers was found to be less than 1\% and polarization
proved to be stable (rotation less than $1^\circ$) within an hour.

Each of the observers switched the direction of local polarization
analysis with a transverse electro-optic modulator. It's optic
axes was set at $45^\circ$ with respect to the subsequent polarizer.
Applying a voltage causes a rotation of the polarization of light
passing through the modulator by a certain angle proportional to
the voltage\cite{footnote1}. For the measurements the modulators
were switched fast between a rotation of $0^\circ$ and $45^\circ$.

The modulation systems (high-voltage amplifier and electro-optic
modulator) had a frequency range from DC to 30~MHz. Operating the
systems at high frequencies we observed a reduced polarization
contrast of 97\% (Bob) and 98\% (Alice). This, however, is no real
depolarization but merely reflects the fact that we are averaging
over the polarization rotation induced by an electrical signal
from the high-voltage amplifier, which is not of perfectly
rectangular shape.

\epsfigure{One of the two observer stations. A random number
generator is driving the electro-optic modulator. Silicon
avalanche photodiodes are used as detectors. A ``time tag'' is
stored for each detected photon together with the corresponding
random number ``0'' or ``1'' and the code for the detector ``$+$''
or ``$-$'' corresponding to the two outputs of the Wollaston prism
polarizer. All alignments and adjustments were pure local
operations that did not rely on a common source or on
communication between the observers.
\label{Observer}}{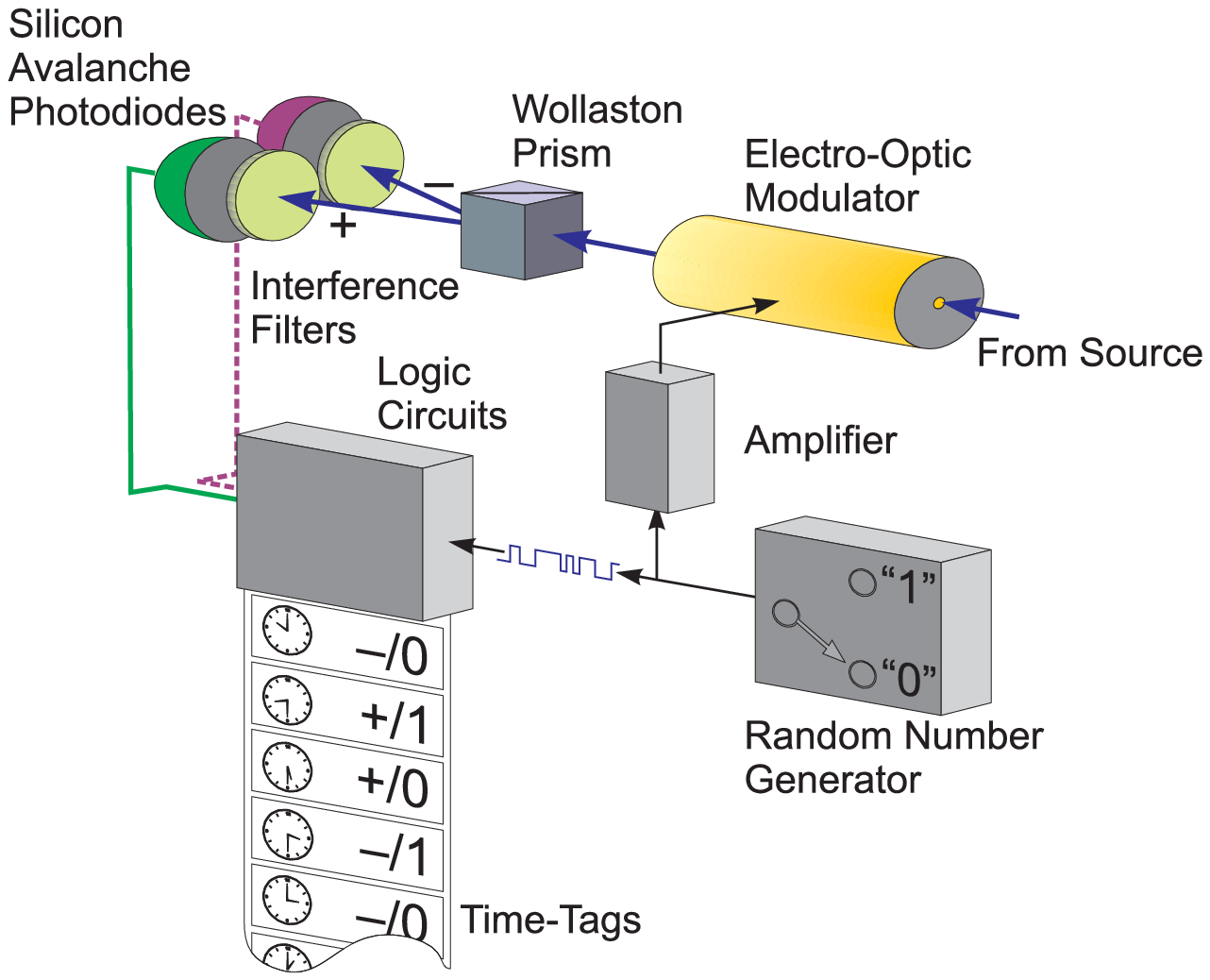}{\columnwidth}

The actual orientation for local polarization analysis was
determined independently by a physical random number generator.
This generator has a light-emitting diode (coherence time $t_c
\approx $10~fs) illuminating a beam-splitter whose outputs are
monitored by photomultipliers. The subsequent electronic circuit
sets its output to ``0''(``1'') upon receiving a pulse from
photomultiplier ``0''(``1''). Events where both photomultipliers
register a photon within $\triangle t\leq$~2~ns are ignored. The
resulting binary random number generator has a maximum toggle
frequency of 500~MHz. By changing the source intensity the mean
interval was adjusted to about 10~ns in order to have a high
primary random bit rate\cite{Achleitner97,footnote2}. Certainly
this kind of random-number generator is not necessarily evenly
distributed. For a test of Bell's inequality it is, however, not
necessary to have perfectly even distribution, because all
correlation functions are normalized to the total number of events
for a certain combination of the analyzers' settings. Still, we
kept the distribution even to within 2\% in order to obtain an
approximately equal number of samples for each setting. The
distribution was adjusted by equalizing the number of counts of
the two photomultipliers through changing their internal
photoelectron amplification. Due to the limited speed of the
subsequent modulation system it was sufficient to sample this
random number generator periodically at a rate of 10~MHz.

The total of the delays occurring in the electronics and optics of
our random number generator, sampling circuit, amplifier,
electro-optic modulator and avalanche photodiodes was measured to
be 75~ns. Allowing for another 25~ns, to be sure that the
autocorrelation of the random number generator output signal is
sufficiently low, it was safe to assume that the specific choice
of an analyzer setting would not be influenced by any event more
than 100~ns earlier. This was much shorter than the $1.3 \mu$s
that any information about the other observer's measurement would
have been retarded.

The photons were detected by silicon avalanche photodiodes with
dark count rates (noise) of a few hundred per second. This is very
small compared to the 10.000 -- 15.000 signal counts per second
per detector. The pulses of each detector were fed into constant
fraction discriminators to achieve accurate timing, and from there
into the logic circuits. These logic circuits were responsible for
disregarding events that occurred during transitions of the switch
signal and generating an extra signal indicating the position of
the switch. Finally all signals were time-tagged in special time
interval analyzers, which allowed us to record the events with
75~ps resolution and 0.5~ns accuracy referenced to a rubidium
standard together with the appendant switch position. The overall
dead time of an individual detection channel was approximately $1
{\mu}s$.

Using an auxiliary input of our time interval analyzers we
synchronized Alice's and Bob's time scales by sending laser pulses
(670~nm wavelength, 3~ns width) through a second optical fiber
from the center to the observing stations. While the actual jitter
between these pulses was measured in the laboratory to be less
than 0.5~ns, the auxiliary input of the time interval analyzers
had a resolution not better than 20~ns thus limiting
synchronization accuracy. This non-perfect synchronization only
limited our ability to exactly predict the apparent time shift
between Alice's and Bob's data series, but did not in any way
degrade time resolution or accuracy. It is important, however,
that this uncertainty was smaller than the dead time of our
detectors, because otherwise data analysis would have been much
more complex.

Each observer station featured a PC which stored the tables of
time tags accumulated in an individual measurement. Long after
measurements were finished we analyzed the files for coincidences
with a third computer. Coincidences were identified by calculating
time differences between Alice's and Bob's time tags and comparing
these with a time window (typically a few ns). As there were four
channels on each side --- two detectors with two switch positions
--- this procedure yielded 16 coincidence rates, appropriate for
the analysis of Bell's inequality. The coincidence peak was nearly
noise-free (${\rm SNR} > 100$) with approximately Gaussian shape
and a width (FWHM) of about 2~ns. All data reported here were
calculated with a window of 6~ns.

There are many variants of Bell's inequalities. Here we use a
version first derived by Clauser et al.\cite{Clauser69} (CHSH)
since it applies directly to our experimental configuration. The
number of coincidences between Alice's detector $i$ and Bob's
detector $j$ is denoted by $C_{ij}(\alpha,\beta)$ with
$i,j\in\{\rm{+,-}\}$ where $\alpha$ and $\beta$ are the directions
of the two polarization analyzers and ``$+$'' and ``$-$'' denote
the two outputs of a two-channel polarizer respectively. If we
assume that the detected pairs are a fair sample of all pairs
emitted, then the normalized expectation value $E(\alpha,\beta)$
of the correlation between Alice's and Bob's local results is $
E(\alpha,\beta) =
[C_{++}(\alpha,\beta)+C_{--}(\alpha,\beta)-
C_{+-}(\alpha,\beta)-C_{-+}(\alpha,\beta)]/N$,
where $N$ is the sum of all coincidence rates. In a rather general
form the CHSH inequality reads

\begin{eqnarray}
 S(\alpha,\alpha',\beta,\beta') & = &
|E(\alpha,\beta)-E(\alpha',\beta)|+ \nonumber \\ & & +
|E(\alpha,\beta')+E(\alpha',\beta')| \leq 2.
\end{eqnarray}

Quantum theory predicts a sinusoidal dependence for the
coincidence rate $C^{qm}_{++}(\alpha,\beta) \propto \sin
^2(\beta-\alpha)$ on the difference angle of the analyzer
directions in Alice's and Bob's experiments. The same behavior can
also be seen in the correlation function $E^{qm}(\alpha,\beta) =
-\cos(2(\beta-\alpha))$. Thus, for various combinations of
analyzer directions $\alpha,\beta,\alpha',\beta'$ these functions
violate Bell's inequality. Maximum violation is obtained using the
following set of angles $S^{qm}_{\rm max}
=S^{qm}(0^\circ,45^\circ,22.5^\circ,67.5^\circ)= 2\sqrt{2} = 2.82>2$

If, however, the perfect correlations ($\alpha-\beta = 0^\circ$ or
$90^\circ$) have a reduced visibility $V \leq 1$ then the quantum
theoretical predictions for $E$ and $S$ are reduced as well by the
same factor independent of the angle. Thus, because the visibility
of the perfect correlations in our experiment was about 97\% we
expect $S$ to be not higher than 2.74 if alignment of all angles
is perfect and all detectors are equally efficient.

We performed various measurements with the described setup. The
data presented in Fig.~\ref{Data} are the result of a scan of the
DC bias voltage in Alice's modulation system over a 200~V range in
5~V steps. At each point a synchronization pulse triggered a
measurement period of 5~s on each side. From the time-tag series
we extracted coincidences after all measurements had been
finished. Fig.~\ref{Data} shows four of the 16 resulting
coincidence rates as functions of the bias voltage. Each curve
corresponds to a certain detector and a certain modulator state on
each side. A nonlinear $\chi^2$-fit showed perfect agreement with
the sine curve predicted by quantum theory. Visibility was 97\% as
one could have expected from the previously measured polarization
contrast. No oscillations in the singles count rates were found.
We want to stress again that the accidental coincidences have not
been subtracted from the plotted data.

In order to give quantitative results for the violation of Bell's
inequality with better statistics, we performed experimental runs
with the settings $0^\circ, 45^\circ$ for Alice's and $22.5^\circ, 67.5^\circ$ for
Bob's polarization analyzer. A typical observed value of the
function $S$ in such a measurement was $S=2.73 \pm 0.02$ for 14700
coincidence events collected in 10~s. This corresponds to a
violation of the CHSH inequality of 30 standard deviations
assuming only statistical errors. If we allow for asymmetries
between the detectors and minor errors of the modulator voltages
our result agrees very well with the quantum theoretical
prediction.

\epsfigure{Four out of sixteen coincidence rates between various
detection channels as functions of bias voltage (analyzer rotation
angle) on Alice's modulator. A$+$1/B$-$0 for example are the
coincidences between Alice's ``$+$'' detector with switch having
been in position ``1'' and Bob's ``$-$'' detector with switch
position ``0''. The difference in height is explained by different
efficiencies of the detectors.
\label{Data}}{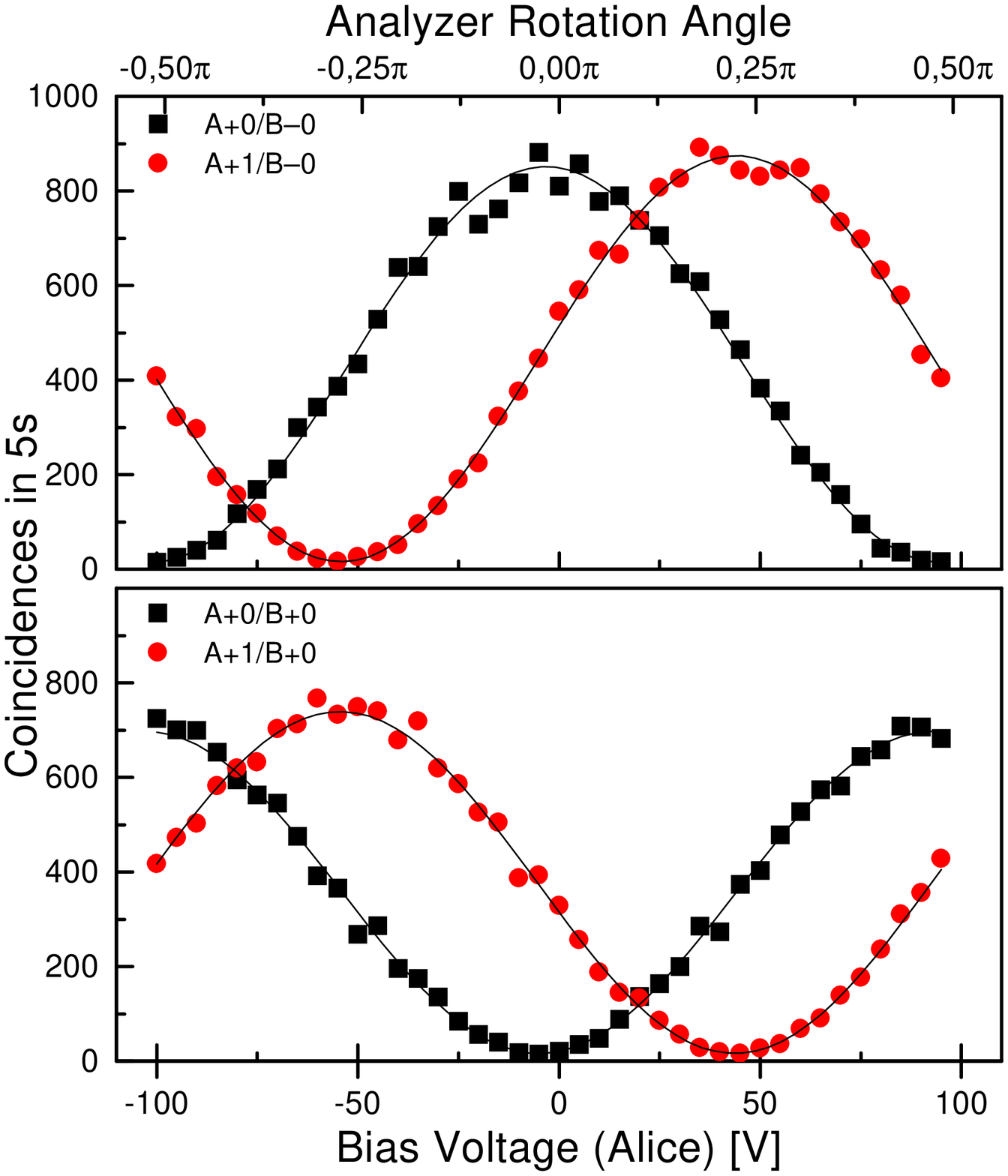}{\columnwidth}

While our results confirm the quantum theoretical
predictions\cite{footnote3}, we admit that, however unlikely,
local realistic or semi-classical interpretations are still
possible. Contrary to all other statistical observations we would
then have to assume that the sample of pairs registered is not a
faithful representative of the whole ensemble emitted. While we
share Bell's judgement about the likelihood of that
explanation\cite{Bell87}, we agree that an ultimate experiment
should also have higher detection/collection efficiency, which was
5\% in our experiment.

Further improvements, e.g. having a human observers choose the
analyzer directions would again necessitate major improvements of
technology as was the case in order to finally, after more than 15
years, go significantly beyond the beautiful 1982 experiment of
Aspect et al\cite{Aspect82}. Expecting that any improved
experiment will also agree with quantum theory, a shift of our
classical philosophical positions seems necessary. Among the
possible implications are nonlocality or complete determinism or
the abandonment of counterfactual conclusions. Whether or not this
will finally answer the eternal question: ``Is the moon there,
when nobody looks?''\cite{Mermin85a}, is certainly up to the
reader's personal judgement.

This work was supported by the Austrian Science Foundation (FWF),
project S6502, by the US NSF grant no. PHY 97-22614, and by the
APART program of the Austrian Academy of Sciences.


\begin{references}
\bibitem{footnote1}
Precisely speaking, the modulator introduces a phase shift between
the linearly polarized components parallel and perpendicular to
its optic axis (at $45^\circ$). Together with two quarter-wave plates
(at $0^\circ$ or $90^\circ$) before and after the modulator this results in
a polarization rotation in real space as usually seen in
circularly birefringent media. The latter quarter-wave plate can
be abandoned here because it is parallel to the axis of the
subsequent polarizer and thus introduces only a phase which cannot
be measured anyway. The quarter-wave plate in front of the
modulator is substituted by our fiber and the initial polarization
controllers.
\bibitem{footnote2} We tested our random-number-generator for
exponential distribution of the occurring time intervals, for
maximum entropy, and for short autocorrelation. Further we applied
many special tests normally used to judge the quality of
pseudo-random-number-generators \cite{Marsaglia95a}. These results
will be the subject of a future publication.
\end{references}
\end{document}